# Solar Gravity Modes : Present and Future

Sylvaine Turck-Chièze, cturck@cea.fr
*DAPNIA,  CE Saclay, CEA, 91191 Gif sur Yvette Cedex, France*

**Abstract**

Gravity modes are the best probes to study the solar radiative zone dynamics, especially in the nuclear core. These modes remain difficult to observe, but they are essential ingredients for progressing on the evolution of the Sun-Earth relationship at the level of centuries. Today, the knowledge of the internal dynamics comes from acoustic modes and concerns mainly the external 2% of the solar mass. Nevertheless, the flat rotation profile of the radiative zone compels physics beyond the standard framework. I summarize different attempts to look for gravity modes and the results obtained after 8 years of observation with the GOLF/SoHO instrument. Some gravity mode candidates (at 1mm/s level) have appeared with more than 98% confidence level as quadruplets or quintuplets. These patterns, if confirmed as gravity modes, may reveal very exciting physics of the solar core. Getting information on rotation and magnetic field in the solar core are real keys to simulate a complete dynamical solar picture. The understanding of the solar dynamics, the precise energetic balance and its temporal evolution necessitate more observations of the radiative zone which constitutes 98% of the Sun by mass. Our expertise in Doppler velocity measurements allows a step further and a new instrumental concept to reach velocities as low as 0.1 mm/s. A prototype will join the Tenerife site in 2006 and a space version is proposed to CNES and ESA as a microsatellite or part of a payload at the L1 Lagrange point.

## 1.     The long trend of the Sun-Earth relationship

   During the last century, one has understood the slow evolution of stars and their source of energy. The deep understanding of the internal dynamics was considered a secondary objective and it has appeared difficult to model without observational facts. Therefore, the relationship between Sun and Earth is today restricted to a rather simple energetic interaction. The Sun produces energy by nuclear interactions, which equilibrates the surface energy haemorrhage. A constant level of solar energy is sent to the Earth at the human time scale. As a consequence, the earth climate evolution mainly follows a pure relative Sun-Earth position with large periodicities (several thousand years) connected to earth orbital parameters.
   But the climate understanding has pointed out the role of volcanoes and these recent years, more and more questions have been addressed on the role of the solar wind and the coronal mass ejections and their interaction with the earth atmosphere. This evolution shows the importance in understanding the deep origin and evolution of the solar activity, and the interaction of solar particles with the earth magnetosphere. The coupled mission SoHO-CLUSTER, the first cornerstone of ESA attacks this problem for the first time. After SoHO/CLUSTER, the ILWS program will continue to progress on different aspects of this problem.
   This review summarizes the present knowledge of the deep solar interior and some directions of investigation for the next decade. Partly thanks to SoHO, it is now established that solar activity has an internal origin, and that the convective zone (2% of the mass) plays an important role.  Moreover, the presence of a thin turbulent layer at its bottom appears essential (Spiegel & Zahn 1992) in understanding the 22 year periodicities (Dikpati & Gilman 2001; Bonanno et al. 2002). In this layer, the observed surface differential rotation suddenly vanishes, given the name of tachocline to this transition zone between convection and radiation transport of energy. The interest for this region reinforces the need to better investigate the radiative zone itself which contains 98% of the solar mass. This region appears nowadays more complex than described up to now and its magnetic field plays a fundamental role in the dynamical structure of the present Sun (Mathis & Zahn 2005). The redistribution of energy due to meridional circulation, kinetic and magnetic energy can be studied (Brun & Zahn 2005 to be published). Gravity modes are the best probes to study the dynamics of the solar radiative zone. Until now, we know nothing on its variability with time and latitude.
    In this review, we summarize, in section 2, the best results obtained on the radiative zone with the SoHO satellite coming from acoustic modes and why they warrent improvements in solar modelling. The strategies adopted to look for small signals in the range of gravity modes and the present results are given in section 3. In section 4, we describe the objectives and the present status of the European projects and the important physical questions which they would like to address. We conclude in section 5.

## 2. Toward a renewal of solar modelling

Acoustic and gravity modes are characterized by three numbers: (1) the degree $\ell$ which corresponds to the number of reflections at the surface: looking at the Sun globally gives access to the most penetrating modes of degree $\ell$ = 0, 1, 2, 3; (2) the order n is the number of nodes along the radius (typically for each degree, we access to 35-40 harmonics n); (3) the component m which provides information on the aspheric sun due to rotation and magnetic field, at each degree $\ell$ corresponds 2m+1 components. Our knowledge of the solar radiative zone has dramatically been improved thanks to the detection and the identification of about 500 different acoustic modes obtained with GOLF and MDI aboard SoHO and the ground networks.

Information on the radiative zone with acoustic modes is in reality difficult to extract because these modes are mainly sensitive to the physics of the outer layers. We have shown that the contribution of the nuclear region to the absolute acoustic frequencies is evidently small (figure 1 of Garcia et al. 2001) but the low-order (low frequencies) modes, below 2 mHz, are not polluted by the solar cycle nor by the stochastic process because the lifetime of these modes is longer and the external turning point lower. Space GOLF/MDI instruments have been used together to detect and confirm such modes after 3 years of observations (Bertello et al. 2000, Garcia et al., 2004b). These modes have been essential to get a precise insight on the solar core (Turck-Chièze et al. 2001, Couvidat et al. 2003a). Their knowledge leads to an "unprecedented accuracy" on the central solar region. The sound speed is now precisely determined (at $10^{-4}$ accuracy) down to 0.06 $R_\odot$. This extended range of detection is especially important for the extraction of the rotation profile in the radiative zone: a very clean rotation extraction is now possible because each splitting (distance between 2 components m) is determined with an uncertainty of about 3%. Consequently, the rotation profile is now clearly established down to the core limit (0.2 $R_\odot$) (Couvidat et al., 2003b; Garcia et al, 2004a) and the flat profile in the radiative zone, in contrast to a differential rotation in the convective zone, can only be explained by invoking the presence of a magnetic field or internal waves. Such information confirms and completes that obtained by BiSON and IRIS networks on longer series (Chaplin et al. 2002, Fletcher et al. 2003, Salabert et al. 2004). All these results show that the global Doppler velocity technique developed in Europe is a powerful one.

Standard models have evolved in parallel including new updated classical physics (Basu et al. 2000, Turck-Chièze et al. 2001a, b, Bahcall et al. 2001, Guzik et al. 2001, Turck-Chièze et al. 2004b, Antia & Basu 2005…). The seismic and neutrino probes have contributed in improving some fundamental ingredients sometimes by a factor 2 ($^{14}$N(p,$\gamma$) reaction rate) or 30% (CNO abundances) or even more (neon case ?). Thus today, the thermodynamic structure of the Sun is reasonably under control, but the existing solar modelling cannot reproduce the rotation profile because neither rotation nor magnetic fields are present in the equations. Nevertheless we know how to make such improvement (Mathis et Zahn 2004, 2005). These last years, we have built a "seismic" model which mimics the observed radiative region sound speed in adjusting values of the standard model physical ingredients (Turck-Chièze et al. 2001b, Couvidat et al. 2003a). This model contains the best physical processes describing the solar plasma: turbulence, relativistic and screening effects, microscopic diffusion …. It is built to show quantitatively the difference between the Sun and the standard model, but it is not a final model as it does not contain the dynamical processes like meridional circulation, magnetic field and dynamo. But the seismic model is the closer model of the Sun today and its main interest is to be a predictable model. It gives a quantitative description of the central plasma characteristics from which one can deduce the central temperature, now determined to better than 1%. It is used to predict gravity mode predictions and the different neutrino fluxes, independently of any doubt on a specific classical ingredient. It is interesting to note their remarkable agreement with the detected neutrinos: they are directly compared to the SNO results and also agree with the other detections after the introduction of neutrino oscillation parameters deduced from the different measured neutrino fluxes (Turck-Chièze et al. 2004b, c).

It is important to note the instabilities of the standard models these last years. After the introduction of recent reaction rates and the updated on CNO composition of the Sun, the standard model disagrees with both seismic observations and solar neutrino detections (Turck-Chièze 2004b; Guzik & Watson 2004). These results show that we are in a transition period, with a clear need for physics beyond the classical physics and possible new improvements in essential ingredients as opacities or specific abundance.

In fact, improvements in solar, and of course stellar modelling have been waiting for a long time. All stars are rotating and consequent stellar internal behaviours have been studied by Mestel (1953) and Schatzman (1962) but without observational constraints the progress has been slow during the last 30 years. The seismic investigation of the Sun today gives important constraints for stellar modelling. The effects are known to be even more important for young stars or very massive stars. In the radiative stellar zones, the rotation induces thermally driven meridional circulation and turbulence generated by the shear of the differential rotation, which is probably mainly anisotropic with strong transport in the horizontal direction. The introduction of this turbulent term in the diffusion of elements of the classical evolution code (Brun, Turck-Chièze, Zahn, 1999) has allowed us to better understand the lithium depletion in main sequence stars, but seems insufficient to describe lithium in premain-sequence (Piau & Turck-Chièze 2002). Moreover it has been demonstrated that the existence of a thin tachocline supposes the existence of a small radial magnetic field, of the order of $10^{-4}$ G with a toroidal field of about 200 G (Rüdiger & Kitchatinov 1997, Gough & McIntyre 1998).

So, magnetic fields in the radiative zone must be added to pure hydrodynamics effects to reproduce the internal phenomena and the solar rotation profile observed with the acoustic modes. Hydrodynamical simulations alone cannot reproduce the non shedular radial rotation profile of the Sun. The difficulty is to find a way to introduce all these new phenomena in stellar evolution codes. Several directions are considered: (1) the improvement of the 1D stellar evolution equations, in doing reasonable hypotheses on the different processes following Maeder & Zahn (1998). This is the only way to follow the stellar evolution on long time scale with the present computer performances. This approach can estimate the total energy budget and consequently the actual long-trend of the Sun at the human timescale. It will also help to discover other magnetic cycles of the Sun. This approach will introduce the effects of rotation, magnetic field and internal gravity waves on the transport of angular momentum. It breaks the apparent symmetry between the circulation driven by the differential rotation and the counteracting flow induced by the non homogeneous chemical composition (Mathis & Zahn 2005). Our objective is to build a general scheme for different stellar masses at different stages of evolution. Stellar seismology with COROT will bring new constraints for other stars and comparison between observations and predictions will largely benefit from such approach, (2) the building of a 2D stellar evolution code: there is a need for this approach because rapid rotation produces strong deformation. Several groups are progressing in this direction. (3) The development of 3D MHD simulations of parts of stars. This approach introduces naturally all the magneto-hydrodynamic processes but they are limited in time steps and resolution, even they show the fluid reactions and put in evidence the different actors of the dynamo. This approach is developing quickly with qualitative information on the energy distribution, the first objective is to reproduce the differential rotation of the solar convective zone (Brun, Miesch, Toomre, 2004). The simulation of the radiative zone and the tachocline is the next step and observations are welcome to guide them.

## 3. Solar gravity mode search aboard SoHO

As mentioned in previous sections, acoustic modes have their maximum of sensitivity in the outer layers, so they are indirect observables of the radiative zone. On the contrary, the maximum sensitivity of the gravity modes is in the solar core. They must be detected if we want to progress on the dynamics of the radiative zone. Gravity modes have been searched for more than 20 years. The limit of detection on ground was 7 cm/s (Delache and Scherrer, 1983; van der Raay 1990) before SoHO. The satellite IPHIR using luminosity variations has not really changed the detection limit: 1.3 ppm at 20 $\mu$Hz corresponds to some cm/s (Frölich et al. 1991). GOLF instrument has been specifically designed to detect velocities down to 1 mm/s thanks to very low instrumental noise (Gabriel et al. 1997, see figure 2 of Turck-Chièze et al., 2004a). Before SoHO launch, all the searches were oriented to patterns spaced in period which are the characteristics of gravity modes below 100 $\mu$Hz. But the theoretical predictions (Andersen 1996, Kumar et al., 1996), just after launch, have shown that the gravity modes in this frequency range, have a visibility at least a factor 10 smaller than those appearing in the upper part of the gravity frequency range. Moreover, they have shown that their surface velocity may be fraction of mm/s. So the solar ground networks BiSON and IRIS seem excluded from an exploration of this region due to atmospheric, instrumental perturbations and duty cycles.

Since the launch of SoHO, several attempts have been dedicated to gravity mode detection. A limit of 1 cm/s for single peaks have been obtained by Appourchaux et al. (2000) looking to MDI, VIRGO, BiSON and GONG network data, then the limit was reduced to 6 mm/s for the GOLF instrument after 4-5 years (Gabriel et al., 2002). In parallel, Turck-Chièze et al. (1998) have developed a strategy to search for gravity mode multiplets and applied it to the GOLF instrument (Gabriel et al. 1998). It consists (1) to look for gravity modes above 150 $\mu$Hz, as their velocities are enhanced in comparison with lower frequencies, (2) to look for multiplets instead of single peaks as they are informative on the central rotation profile and allow a lower detection limit more compatible with theoretical predictions, (3) to use different analysis techniques. Such patterns have then been searched automatically with the criterion to have more than 90% confidence level not to be pure noise and followed in time to eliminate evident noise candidates (Turck-Chièze et al. 2004a). We have also used our knowledge of the Sun given by acoustic mode detection through the seismic model to eliminate unphysical candidates and to compare with theoretical predictions (Provost et al. 2000). After 8 years of observations (2975 days for GOLF and 2925 days for MDI), the solar granulation noise limits our detection capability at frequencies lower than 1 mHz but the lowest order acoustic modes (typically n= 1 or 3) appear easier to access due to their mixed character. In the region 150-450 μHz, one "gravity mode candidate" is detected within 10 μHz around the theoretical value of the seismic model, with more than 98% confidence level as a quadruplet or a quintuplet. The same analysis on MDI data shows several peaks of the pattern. 4 peaks are observed in GOLF data since the first 800 days analysis and a new analysis of the first 1290 days shows that it was detected also, as a quadruplet, at the same confidence level. A second case around 265 μHz is detected with more than 90% confidence level, other patterns previously mentioned as quintuplets at 90% have completely disappeared (Turck-Chièze et al., 2004d). Of course these patterns must be pure noise. But the persistence of more than 2 or 3 peaks at a position where we are waiting some $\ell = 2$ or $\ell = 1$ with a presence of $\ell = 5$, or 4 in the neighbour is interesting to notice.

The solar granulation noise is the dominant contribution to the signal in the corresponding range of frequencies for GOLF (figure 2 of Turck-Chièze, 2004a). Thus long observations help the detection of such faint gravity mode signals, even the GOLF instrument has observed in a restricted mode (Ulrich et al. 2000, Garcia et al. 2005) if the instrumental noise does not evolve too much in time and if the pattern is not too complex.

On the theoretical side, Dintrans et al. (2005) show that these gravity modes probably have a reduced lifetime of the order of twice the period, and could appear mainly episodically. Moreover, Cox and Guzik (2004) have remarked that the visibility of the specific mode $\ell = 2$, n = -3 is predicted to be larger than the neighbour ones, so they suggest that the persistent observation made with GOLF could reasonably be associated to this mode.

If this persistent detection is not pure noise, interesting physical questions are associated with the observed pattern: the "detected" peaks do not appear extremely thin and show some structure (see figures in Turck-Chièze 2004a,d) as were mentioned several years ago by Goode and Thompson (1992). A central magnetic field can spread the power and introduces a hyperfine structure which is interesting to detect (Turck-Chièze et al. 2005a). Moreover, if one detects a quintuplet instead of a triplet, such detection could reveal a different rotation axis for the solar core than for the rest of the radiative zone, perhaps a relic of the young sun. Moreover the observed spacing would be in favour of a core rotating more rapidly due to a splitting of about 0.6 μHz (instead about 0.4 μHz) than the rest of the radiative zone. All these possible phenomena push the observers to pursue and improve the gravity mode detection while there is still room for less exciting explanations as only interplay between modes or pure noise. The observations will continue up to the end of the mission of SoHO but the natural degradation of GOLF observations does not give important hope of significant improvement.

Therefore we are in parallel preparing a next generation of instruments capable to pursue this search during the observations of the SDO mission, with complementary instruments. We have effectively demonstrated the interest in using two good instruments for the best extraction of the acoustic mode science.

## 4. Future gravity mode detections

Today we are facing two problems which limit the gravity mode detection: (1) their extremely small surface velocity, (2) the presence of solar noise in the region which contributes to poor signal/ noise.
The next generation of instruments dedicated to gravity modes must address these points to observe quicker (1-2 years instead of typically 5 years in space, about 10 years on ground) and to observe with a better sensitivity. There is no space instrument which has really been built to face these two questions together. Progress in this range of frequency is not expected to come from the instrument HMI on SDO. Their first objective is the improvement of the spatial resolution to describe with great details the convective zone. The French SODISM instrument, aboard the microsatellite PICARD will benefit by the enhancement of the signal near the solar limb in intensity but the related noise will remain a real problem.

Espagnet et al. (1995) have shown that the Doppler velocity granulation noise is not the same at different heights in the atmosphere of the sodium line. This point has been confirmed, using the first month of observation of the GOLF 4 points instrument: the noise coming from the two wings of the sodium lines separated by 2 km/s present a coherence of only 30%, leading to a potential reduction of the solar noise by a factor 4-5 below 1 mHz when 4 points are used on the sodium line (Garcia et al., 2004c). So in order to improve the signal/noise, we propose to extract the velocity at 7 or 8 heights of the atmosphere. A prototype using this technique is described in Turck-Chièze (2005b), it uses the Doppler velocity method based on a resonant spectrometer as in BiSON and IRIS networks or GOLF aboard SoHO. The detection will be done by 15 channels along the sodium line (8 on each wing with one common at the bottom of the line), the main objective is to contribute to get a MHD picture of the Sun and stars in improving our knowledge of the solar radiative core. The prototype GOLF-NG is presently built by a French Spanish collaboration to solve the technical difficulties and will be put in Tenerife in 2006 to help the SoHO observations at low frequencies.

The improvements for the space version are the following: (1) a lower instrumental detection in improving the detection accuracy. A mean counting rate at $10^8$ cts/s (between $5\ 10^7$ and $5\ 10^8$) gives an accuracy of $10^{-4}$ with a detector noise lower than the statistical noise. (2) a lower solar noise by measuring the velocity at 8 different heights in the atmosphere thanks to a permanent and variable linear spectrometer between 0 to 8 kG. The sodium line is particularly adapted for this improvement as the slope of its wings is not abrupt, (3) a measurement of the continuum near the sodium line in order to properly differentiate intensity from velocity variations and shifts of the line due to movements of the solar atmosphere, this last point may contribute to reduce other sources of solar noise during periods of increased activity, (4) some minimal macro pixels to detect gravity modes up to $\ell = 5$ in order to be able to identify the observed modes by different masks. This instrument, called DynaMICS (Dynamics and Magnetism of the Internal Core of the Sun) will be a useful companion for the two previously mentioned instruments launched in 2008. This instrument will continue the exploration of the low frequency acoustic modes and explore the region of mixed modes which contain a very rich physics. It will also contribute to follow in time the evolution of the modes with the solar cycle and explore the region of chromospheric modes at high frequencies, it will put more constraints on the mean behaviour of the 600 km above the photosphere. This instrument is an important element for the near future so a space version must be developed quickly for a microsatellite and (or) be part of the payload of the sentinel sent at the L1 Lagrange point in the context of the ILWS mission.

## 5. Conclusion and Perspectives

SoHO instruments have improved our knowledge of the solar radiative zone by the detection of a substantial number of acoustic mode frequencies below 2 mHz which are not polluted by stochastic excitation or magnetic variations of the outer layers. The mean behaviour of this important solar zone is now properly described, due to the combinaison of the velocity Doppler technique, the low instrumental noise and the long duration of SoHO observations which allows detection of signal down to 1-2 mm/s.

The next step is the unambiguous detection of gravity modes, which are the only real probes of the core dynamics. The present efforts are devoted to the transition region between gravity and acoustic modes (150 $\mu$Hz-1 mHz). Specific patterns of quadruplets or quintuplets are visible in the GOLF instrument at high confidence level (98%). They decrease the detection limit for gravity modes by a factor 40. The knowledge of only several gravity modes will reveal the main characteristics of the solar core dynamics which is announced to be rather complex: high rotation ? high magnetic field ? differential rotation ?

The detection of low frequency modes by the resonant scattering spectrophotometer technique has always been successful, first on ground, then in space. We know how to improve the performances of this technique by at least a factor 5 to 10. This is necessary to guide simulations of a 3D MHD solar picture including the estimate of the real energetic balance. It is why a new generation of instruments is presently in construction to identify the characteristics of the first gravity modes during the coming decade and to better determine magnetic field and rotation in the radiative zone. This progress is extremely useful for the prediction of the long-term magnetic behaviour of the Sun and for the development of asteroseismology.

It is important to determine and quantify the different sources (if they are several) of the solar activity, and their impact on the Earth evolution. The first step will be in Europe, the launch of PICARD, in 2008 which will quantify the radius deformation and the luminosity variations in different wavelengths.

Europe will gain to include DynaMICS in complement to PICARD, in the ILWS program of the coming solar cycle. It uses a well known technique and will lower the solar granulation noise. This decision will guarantee the best scientific return of the coming decade, in putting in action all the existing successful techniques, without excluding the one which has provided constant progress and shown its ability to be improved by a substantial amount. This step is necessary to prepare the world mission of the ESA Cosmic Vision and to improve our knowledge of magnetic fields inside the Sun (Turck-Chièze et al. 2005a). For the far future, other ideas are being studied, such as measuring gravity waves arising from the Sun (Appourchaux 2003).

## 6. Acknowledgement

I would like to thank the whole helioseismic community together with the CNES, CEA and IAC institutions for their support in the analysis of the GOLF instrument and the building of the GOLF-NG prototype.